\documentclass[12pt,english]{article}
\usepackage{babel}
\usepackage{psfrag,graphicx,float}
\usepackage{amsmath,amssymb}
\usepackage[mathscr]{eucal}

\def\be{\begin{equation}}
\def\ee{\end{equation}}
\def\ba{\begin{eqnarray}}
\def\ea{\end{eqnarray}}
\def\bs{\begin{subequations}}
\def\es{\end{subequations}}

\begin{document}

\title{The Kundu--Eckhaus equation and  its discretizations}
\author{ Decio Levi$^1$ and   Christian Scimiterna\\
\\  Dipartimento di Ingegneria Elettronica \\ Universit\`a degli Studi Roma Tre \\ Via della Vasca Navale 84, 00146 Roma, Italy\\ levi@roma3.infn.it, scimiterna@fis.uniroma3.it\\
$^1$INFN Sezione di Roma Tre}

\maketitle
\begin{abstract}
In this article we show that the complex Burgers and the Kundu--Eckhaus equations are related by a Miura transformation. We use this relation to discretize the Kundu--Eckhaus equation.
\end{abstract}

\section{Introduction}
The Burgers equations 
is the simplest partial differential equation that combines
nonlinear wave propagation with diffusive effects.
As such, it has been widely applied to the modelling
of physical processes such as sedimentation, shock
propagation in gaseous flow, turbulence in fluids and
road traffic \cite{burgers,whitham}.  The name of Burgers equation was introduced by Hopf as a reference to the results of Burgers \cite{burgers} but, in fact, the Burgers equation  can be found already  in
 a work by Bateman published in 1915 \cite{bateman}.  An explicit solution of the Cauchy problem on the
infinite line for the Burgers equation
may be obtained by means of the linearizing
Hopf-Cole transform, introduced indepedently by Hopf and Cole in 1950 \cite{cole,hopf}. This transformation is already contained in an article by Florin \cite{florin} published in 1948 and implicitly 
in a book by Forsyth \cite{forsyth} published in 1906. 

Kundu \cite{kundu} and Eckhaus\cite{CE,E} independently derived in 1984--1985 what can now be called the Kundu--Eckhaus equation as a linearizable form of the Nonlinear Schr\"odinger equation.

Here in the following we will show that the Kundu--Eckhaus equation can be related to the complex Burgers equation by a Miura transformation. Then taking into account this relation we are able to discretize the Kundu--Eckhaus equation using the discretization procedure introduced for the Burgers equation.

In Section \ref{sec2} we present the linearization procedure for both the Burgers and the Kundu--Eckhaus equation and use it to derive the Miura transformation which relate them.  Then in Section \ref{sec3} we use the standard discretization via B\"acklund transformation,  used to construct the discrete Burgers equation, to discretize the Kundu--Eckhaus equation. 

\section{The complex Burgers equation and the Kundu--Eckhaus equation} \label{sec2}

 Let us consider a complex extension of the Burgers 
\ba \label{a1}
i u_t + u_{xx} + 2 u u_x =0
\ea 
where $u(x,t)$ is a complex field. Eq. (\ref{a1}), introducing  the standard Hopf--Cole transformation
\ba \label{a2}
u(x,t) = \frac{\phi_x}{\phi},
\ea 
reduces to the time dependent free Schr\"odinger equation
\ba \label{a3}
i \phi_t + \phi_{xx} = 0,
\ea 
provided that the time evolution of the function $\phi(x,t)$  satisfies the linear ordinary differential equation
\ba
 \phi_t= i(u_{x}+u^2)\phi\Bigl |_{x=a},\label{a11}
\ea
where $a$ is an arbitrary value of the $x$ variable at which all the functions involved are well defined. The solution of eq. (\ref{a11}) gives
\ba
\phi(a,t)=\phi(a,b)e^{i\int_{b}^{t}(u_{x}+u^2)\vert_{x=a}dt^{\prime}},\label{a111}
\ea
where $b$ is an arbitrary  value of the $t$ variable at which all the functions involved are well defined.. Consequently, as it is well known,  the inverse of the Hopf--Cole transformation (\ref{a2}) reads
\ba
\phi(x,t)=\phi(a,t)e^{\int_{a}^{x}u dx^{\prime}}.\label{a1111}
\ea

 A less known linearizable equation is the Kundu--Eckhaus equation
\ba \label{a4}
i \psi_t + \psi_{xx} + 2 \psi |\psi|^2_x + \psi |\psi|^4 = 0.
\ea 
This is a nonlinear Schr\"odinger type equation that also linearizes to the free linear Schr\"odinger
equation (\ref{a3}). As the well known Nonlinear Schr\"odinger equation, it is a universal model equation and, as such, it  appears in many applications. For example it has been  obtained  in the
study of the instabilities of plane solitons associated
with the Kadomtsev-Petviashvili equation \cite{2}. 

The Kundu--Eckhaus equation is linearizable to the time dependent free Schr\"odinger equation (\ref{a3}) through the following procedure. 
Let us define the complex function 
\ba \label{a5}
\phi = \sqrt{2 \Phi} \psi,
\ea 
where the real function $\Phi$ is related to $\phi$ by the following overdetermined system of equations
\begin{subequations} \label{a6}
\ba  
\Phi_x &=& |\phi|^2,\label{a6a}\\ 
\Phi_t &=& i [ \bar \phi \phi_x - \phi \bar \phi_x ],\label{a6b}
\ea
\end{subequations}
where by a bar we indicate the complex conjugate. Inserting eq. (\ref{a5}) into eq. (\ref{a3}) and taking into account eqs. (\ref{a6}), we get the Kundu--Eckhaus equation (\ref{a4}). The compatibility of eqs. (\ref{a6}) is identically satisfied on the solutions of eq. (\ref{a3}). Solving eqs. (\ref{a6}), we get
\ba \label{aa1}
\Phi &=& \int_{a}^{x}|\phi|^2dx^{\prime}+\frac{1}{2} \rho(t), \\ \label{a7b} \rho(t)&=&2i\int_{b}^{t}\left(\bar\phi\phi_{x}-\phi\bar\phi_{x}\right)\vert_{x=a}dt^{\prime}+\rho_{0},
\ea
where $\rho_{0}$ is an arbitrary real constant. 
Then 
eq. (\ref{a5}) can be written as
\ba
\psi=\frac{\phi}{\left[2\int_{a}^{x}|\phi|^2dx^{\prime}+\rho(t)\right]^{1/2}},\label{a7a}
\ea
which can be inverted by expressing the function $\Phi$ in eq. (\ref{aa1}) in term of $\psi$ 
\ba
\Phi=\frac{\rho_{0}} {2}e^{2i\int_{b}^{t}\left(\bar\psi\psi_{x}-\psi\bar\psi_{x}\right)\vert_{x=a}dt^{\prime}}e^{2\int_{a}^{x}|\psi|^2dx^{\prime}}.\label{a8b}
\ea
Moreover, as a direct consequence of eq. (\ref{a5}) we get the following relation between $\phi$ and $\psi$ 
\ba\label{a8}
\phi \bar \psi = \bar \phi \psi.
\ea 
By differentiating eq. (\ref{a5}) with respect to $x$ and using eqs. (\ref{a6a}, \ref{a8}) we get the following differential equation relating $\phi$ and $\psi$
\ba \label{a9}
\phi_x = \bigl ( \frac{\psi_x}{\psi} + |\psi|^2 \bigr ) \phi.
\ea
By comparing eqs. (\ref{a2}, \ref{a9}) we obtain a Miura transformation between  the function $\psi$ and  the  function $u$ satisfying the complex Burgers equation (\ref{a1})
\ba \label{a10}
u=\frac{\psi_x}{\psi} + |\psi|^2.\label{a11111}
\ea
The inversion of eq. (\ref{a11111}) is obtained combining eqs. (\ref{a111}, \ref{a1111}, \ref{a7a}). It reads 
\begin{subequations} \label{b11a}
\ba
&&\psi=\frac{A(t)e^{\int_{a}^{x}udx^{\prime}}} {\left[2|A(t)|^2\int_{a}^{x}e^{\int_{a}^{x^{\prime}\left(u+\bar u\right)dx^{\prime\prime}}}dx^{\prime}+\rho(t)/\rho_{0}\right]^{1/2}},\label{b11}\\
&&\rho(t)=2i\rho_{0}\int_{b}^{t}|A(t)|^2\left(u-\bar u\right)\Bigl |_{x=a}dt^{\prime}+\rho_{0},\label{b111}\\
&&A(t)=\psi(a,b)e^{i\int_{b}^{t}(u_{x}+u^2)\vert_{x=a}dt^{\prime}},\label{b1111}
\ea
\end{subequations}
where $\psi(a,b)=\phi(a,b)/\rho_{0}^{1/2}$. It is easy to show that  the Kundu--Eckhaus equation can also be obtained by introducing the Miura transformation (\ref{a10}) into the complex Burgers equation (\ref{a1}) and fixing $\psi(a,t)$ in such a way that it is consistent with eqs. (\ref{b11a}).

This constructive procedure to get the Kundu--Eckhaus equation can be discretized and provide the differential difference and difference difference Kundu--Eckhaus equation.

\section{Discretizations} \label{sec3}

Eq. (\ref{a5}) is a functional relation and, as such, it is valid when all involved fields, $\Phi$, $\phi$ and $\psi$, depend not only on continuous variables but also  on discrete variables. Similarly, as a consequence, the same is true for eq. (\ref{a8}). What will change when discretizing is the linear equation (\ref{a3}), the overdetermined linear system for $\Phi$  and the resulting Burgers and Kundu--Eckhaus equations.

Let us start from the differential difference case when we just discretize the space variable $x$. In this case we assume as a free linear Schr\"odinger equation the differential difference equation
\ba \label{b1}
i \dot \phi_n + \frac{\phi_{n+1} + \phi_{n-1} -2 \phi_n}{h^2} = 0.
\ea 
where $h$ is the lattice spacing and $n$ the lattice index such that $x=nh$. By  compatibility of eq. (\ref{b1}) with the discrete Hopf-Cole transformation
\ba
\frac{\phi_{n+1}-\phi_{n}} {h}=u_{n}\phi_{n},\label{b11s}
\ea
we get the discrete complex Burgers
\ba \label{b5}
i \dot u_n + \frac{ u_{n+1} - 2 u_n+  u_{n-1} }{h^2}  + \frac{1}{h} \bigl [ u_n \left( u_{n+1} - u_n \right) - \frac{  u_{n-1} (u_n - u_{n-1})}{\left(1+hu_{n-1}\right)} \bigr ] = 0.
\ea
The inverse  of  the discrete Hopf-Cole transformation (\ref{b11s}) reads
\begin{subequations}\label{d}
\ba
\phi_{n}=\phi_{a}\prod_{j=a}^{j=n-1}\left(1+hu_{j}\right),\ \ \ n\geq a+1,\\
\phi_{n}=\frac{\phi_{a}}{\prod_{j=n}^{j=a-1}\left(1+hu_{j}\right)},\ \ \ n\leq a-1,
\ea
\end{subequations}
where $\phi_{a}=\phi_{a}(t)$ is  the function $\phi_{n}$ calculated at the arbitrary point $n=a$. When $u_{n}(t)$ satisfies the complex Burgers equation (\ref{b5}),  $\phi_{n}$, given by eq. (\ref{d}), will satisfy the discrete linear Schr\"odinger equation (\ref{b1}) if $\phi_{a}$  satisfies the ordinary differential equation
\ba
i\dot\phi_{n}+\frac{1} {h^2}\left[hu_{n}-1+\frac{1} {\left(1+hu_{n-1}\right)}\right]\phi_{n}\Bigl |_{n=a}=0.\label{d11}
\ea
The solution of eq. (\ref{d11}) is
\ba
\phi_{a}(t)=\phi_{a}(b)e^{\frac{i} {h^2}\int_{b}^{t}\left[hu_{n}-1+\frac{1} {\left(1+hu_{n-1}\right)}\right]\Bigl |_{n=a}dt^{\prime}},\label{d111}
\ea
where $b$ is an arbitrary value of the time variable. 

To construct the differential difference Kundu--Eckhaus equation we replace the overdetermined system of equations for $\Phi(x,t)$ by a system for $\Phi_n(t)$, whose compatibility is satisfied on the solutions of the differential difference linear Schr\"odinger equation (\ref{d11}). We get 
\bs \label{b20}
\ba \label{b2}
&&\Phi_{n+1} - \Phi_n = h |\phi_n|^2, \\ \label{b3}
&&\dot \Phi_n = \frac{i}{h} \bigl ( \bar \phi_{n-1} \phi_n - \bar \phi_n \phi_{n-1} \bigr ).
\ea 
\es
Solving eqs. (\ref{b20}), we get that eq. (\ref{a5}) becomes
\begin{subequations}\label{b70}
\ba
&&\psi_{n}=\frac{\phi_{n}}{\left[2h\sum_{j=a}^{j=n-1}|\phi_{j}|^2+\rho(t)\right]^{1/2}},\ \ \ n\geq a+1,\ \ \label{b7a}\\
&&\psi_{a}=\frac{\phi_{a}}{\sqrt{\rho(t)}},\ \ \ \ \ \ \ \ \ \ \ \ \ \ \ \ \ \ \ \ \ \ \ \ \ \ \ \ \ \ \ \ \ \ \ \ \ \ \ \ \ \ \ \ \ \,\label{b7b}\\
&&\psi_{n}=\frac{\phi_{n}}{\left[-2h\sum_{j=n}^{j=a-1}|\phi_{j}|^2+\rho(t)\right]^{1/2}},\ \ \ n\leq a-1,\label{b7c}\\
&&\rho(t)=\frac{2i} {h}\int_{b}^{t}\bigl ( \bar \phi_{n-1} \phi_n - \bar \phi_n \phi_{n-1} \bigr )\Bigl |_{n=a}dt^{\prime}+\rho_{0},\ \ \ \ \ \ \ \ \ \label{b7d}
\ea
\end{subequations}
where $\rho_{0}$ is an arbitrary real constant. Eqs. (\ref{b70}) can be inverted giving
\begin{subequations}
\ba
&&\phi_{n}=\psi_{n}\sqrt{\rho(t)}\prod_{j=a}^{j=n-1}\left(1+2h|\psi_{j}|^2\right)^{1/2},\ \ \ n\geq a+1,\\
&&\phi_{a}=\psi_{a}\sqrt{\rho(t)},\ \ \ \ \ \ \ \ \ \ \ \ \ \ \ \ \ \ \ \ \ \ \ \ \ \ \ \ \ \ \ \ \ \ \ \ \ \ \ \ \ \ \ \ \ \ \ \,\\
&&\phi_{n}=\frac{\psi_{n}\sqrt{\rho(t)}}{\prod_{j=n}^{j=a-1}\left(1+2h|\psi_{j}|^2\right)^{1/2}},\ \ \ n\leq a-1,\\
&&\rho(t)=\rho_{0}e^{\frac{2i} {h}\int_{b}^{t}\frac{\left(\bar\psi_{n-1}\psi_n -\bar\psi_n\psi_{n-1}\right)} {\left(1+2h|\psi_{n-1}|^2\right)^{1/2}}\Bigl |_{n=a}dt^{\prime}}.\ \ \ \ \ \ \ \ \ \ \ \ \ \ \  \ \ \ \ \ \ \ \, 
\ea
\end{subequations}
From eq. (\ref{a5}), taking into account eqs. (\ref{a8}, \ref{b2}) we get 
\ba \label{b4a}
\phi_{n+1} = \bigl (\frac{\psi_{n+1}}{\psi_n}\sqrt{1+2h |\psi_n|^2}\bigr ) \phi_n.
\ea
By comparing eqs. (\ref{b11s}) and (\ref{b4a}) we get
\ba\label{b4}
u_n = \frac{\psi_{n+1}\sqrt{1+2h |\psi_n|^2}-\psi_{n}}{h\psi_{n}},
\ea 
that is the discrete Miura transformation between the function $\psi_{n}(t)$ and the function $u_{n}(t)$ following the complex differential difference Burgers equation (\ref{b5}). The inversion of eq. (\ref{b4}) is obtained considering eqs. (\ref{d}, \ref{d111}, \ref{b70}), getting
\begin{subequations} \label{b71}
\ba
&&\psi_{n}=\frac{A(t)\prod_{j=a}^{j=n-1}\left(1+hu_{j}\right)} {\left[2h|A(t)|^2\sum_{j=a}^{j=n-1}\prod_{k=a}^{k=j-1}|1+hu_{k}|^2+\rho(t)/\rho_{0}\right]^{1/2}},\ \ \ \ \ \label{e1}\\
&&\psi_{a}=\frac{A(t)} {\sqrt{\rho(t)/\rho_{0}}},\ \ \ \ \ \ \ \ \ \ \ \ \ \ \ \ \ \ \ \ \ \ \ \ \ \ \ \ \ \ \ \ \ \ \ \ \ \ \ \ \ \ \ \ \ \ \ \ \ \ \ \ \ \ \ \ \ \ \ \label{e11}\\
&&\psi_{n}=\frac{A(t)/\prod_{j=n}^{j=a-1}\left(1+hu_{j}\right)} {\left[-2h|A(t)|^2\sum_{j=n}^{j=a-1}1/\prod_{k=j}^{k=a-1}|1+hu_{k}|^2+\rho(t)/\rho_{0}\right]^{1/2}},\label{e111}\\
&&\rho(t)=2i\rho_{0}\int_{b}^{t}|\frac{A(t)} {1+hu_{n-1}}|^2\left(u_{n-1}-\bar u_{n-1}\right)\Bigl |_{n=a}dt^{\prime}+\rho_{0},\ \ \ \ \ \ \ \ \ \label{e1111}\\
&&A(t)=\psi_{a}(b)e^{\frac{i} {h^2}\int_{b}^{t}\left[hu_{n}-1+\frac{1} {\left(1+hu_{n-1}\right)}\right]\Bigl |_{n=a}dt^{\prime}},\ \ \ \ \ \ \ \ \ \ \ \ \ \ \ \ \ \ \ \ \ \ \ \ \ \ \ \ \label{e11111}
\ea
\end{subequations}
where we set $\psi_{a}(b)=\phi_{a}(b)/\rho_{0}^{1/2}$. Eq. (\ref{e1}) is valid for $n \ge a+1$ while eq. (\ref{e111}) for $n \le a+1$.  If one substitutes eq. (\ref{b4}) into the complex Burgers equation (\ref{b5}) fixing $\psi_{a}(t)$ in a consistent way with eqs. (\ref{b71}), we get the following differential difference equation for the function $\psi_n(t)$ 
\ba \label{b5a}
i \dot \psi_n &+& \frac{1}{h^2} \bigl [ \psi_{n+1} \sqrt{1+2h |\psi_n|^2}+ \frac{\psi_{n-1}}{\sqrt{1+2h |\psi_{n-1}|^2}} - 2 \psi_n \bigr ] \\ \nonumber
&-& \frac{1}{h} \frac{\psi_n}{\sqrt{1+2h |\psi_{n-1}|^2}} \bigl [\psi_n \bar \psi_{n-1} - \bar \psi_n \psi_{n-1} \bigr ] = 0, 
\ea 
{\it the differential difference Kundu--Eckhaus equation}. Eq. (\ref{b5a}) can also be obtained by inserting eq. (\ref{a5}) into eq. (\ref{b1}) and   taking into account eqs. (\ref{b20}). Carrying out the continuous limit, when $h \rightarrow 0$ and $n \rightarrow \infty$ in such a way that $x=nh$ remain finite, we recover the Kundu--Eckhaus equation (\ref{a4}).

The completely discrete equation is obtained by discretizing also the time variable and introducing a new index $m$ and its  spacing $\tau$ such that $t=m \tau$. 
In this case, if we want to preserve the linearity of the Lax pair we have to introduce the following  overdetermined system of equations for the real function $\Phi_{n,m}$  
\ba \label{c1}
\Phi_{n+1,m} - \Phi_{n,m} &=& h | \phi_{n,m} |^2, \\ \nonumber
\Phi_{n,m+1} - \Phi_{n,m} &=& \frac{i \tau}{h} \bigl ( \phi_{n,m} \bar  \phi_{n-1,m} - \bar  \phi_{n,m}  \phi_{n-1,m} \bigr )+ \tau \sigma_{n,m}.
\ea 
where $\sigma_{n,m}$ is a real function which goes to zero when $\tau \rightarrow 0$. The compatibility condition of eqs. (\ref{c1}) provide the discrete Sch\"odinger equation 
\ba \label{c2}
\frac{i}{\tau} \bigl ( \phi_{n,m+1} - \phi_{n,m} \bigr ) + \frac{1}{h^2} \bigl ( 
\phi_{n+1,m} + \phi_{n-1,m} - 2 \phi_{n,m} \bigr ) =0,
\ea
if the real function $\sigma_{n,m}$ satisfies the following difference equation
\ba \label{c3}
\sigma_{n+1,m} - \sigma_{n,m} = \frac{\tau}{h^2} \bigl |\phi_{n+1,m} + \phi_{n-1,m} - 2 \phi_{n,m} \bigr |^2
\ea
From eqs. (\ref{a5}, \ref{c1}) we get the following $n$ and $m$ evolution of the function $\phi_{n,m}$
\ba \label{c4}
\phi_{n+1,m} &=& \bigl [ \frac{\psi_{n+1,m}}{\psi_{n,m}} \sqrt{1 + 2 h |\psi_{n,m}|^2} \bigr ] \phi_{n,m}, \\ \nonumber
\phi_{n,m+1} &=& \Bigl [ \frac{\psi_{n,m+1}}{\psi_{n,m}} \sqrt{1 + 2 i\frac{\tau}{h^2} \frac{\psi_{n,m} \bar \psi_{n-1,m}-\psi_{n-1,m} \bar \psi_{n,m} }{\sqrt{1 + 2 h |\psi_{n-1,m}|^2}}+\tau \rho_{n,m} } \Bigr ] \phi_{n,m},
\ea 
where $\rho_{n,m}=\frac{\sigma_{n,m}}{\Phi_{n,m}}$,  taking into account the definition of the function $\sigma_{n,m}$ given by eq. (\ref{c3}), satisfies the first order linear difference equation
 \ba \nonumber
 &&\rho_{n+1,m} - \frac{1}{1 + 2 h |\psi_{n,m}|^2}  \rho_{n,m}=  \frac{2 \tau}{h^3[1 + 2 h |\psi_{n,m}|^2]} \Bigl | \psi_{n+1,m} \sqrt{1+2h |\psi_{n,m}|^2}  \\ \label{c6} &&\qquad \qquad + \frac{\psi_{n-1,m}}{\sqrt{1+2h |\psi_{n-1,m}|^2}} - 2 \psi_{n,m} \Bigr |^2 \Bigr ].
 \ea
Introducing the two relations (\ref{c4}) into the discrete heat equation (\ref{c2})  we get the following nonlinear discrete partial difference equation
\ba \label{c5}
&& i \Bigl [ \psi_{n,m+1} \sqrt{1 + 2 i\frac{\tau}{h^2} \frac{\psi_{n,m} \bar \psi_{n-1,m}-\psi_{n-1,m} \bar \psi_{n,m} }{\sqrt{1 + 2 h |\psi_{n-1,m}|^2}}+\tau \rho_{n,m}} - \psi_{n,m} \Bigr ]   \\ \nonumber &&\qquad +\frac{\tau}{h^2} \bigl [ \psi_{n+1,m} \sqrt{1+2h |\psi_{n,m}|^2}+ \frac{\psi_{n-1,m}}{\sqrt{1+2h |\psi_{n-1,m}|^2}} - 2 \psi_{n,m} \bigr ] =0,
 \ea
 {\it the difference difference Kundu--Eckhaus equation}.
  It is easy to see that in the continuous limit eq. (\ref{c5}) goes into the Kundu--Eckhaus equation (\ref{a4}) independently from the continuous limit of the function $\rho_{n,m}$. Eq. (\ref{c6}) reduces in the same limit to the linear equation $\rho_x = -2 |\psi|^2 \rho + 2 |\psi|^{10}$. Moreover, it is worthwhile to notice that this completely discrete Kundu--Eckhaus equation, as it is  for the well know Nonlinear Schr\"odinger Equation written down by Ablowitz and Ladik \cite{al}, is nonlocal and it involves the function $\psi_{n,m}$ in all points of the lattice. 






\end{document}